\documentclass[conference,a4paper]{IEEEtran}
\IEEEoverridecommandlockouts
\usepackage{cite}
\usepackage{amsmath,amssymb,amsfonts}
\usepackage{graphicx}
\usepackage{textcomp}
\usepackage{xcolor}
\usepackage{comment}
\usepackage{amsmath,amssymb,amsfonts}
\usepackage{tikz}
\usetikzlibrary{arrows,automata,positioning}
\usetikzlibrary{arrows.meta}
\usepackage{graphicx}
\usepackage{textcomp}
\usepackage{xcolor}
\usepackage{comment}
\usepackage{lipsum}
\usepackage{booktabs}
\usepackage{color, colortbl}
\def\BibTeX{{\rm B\kern-.05em{\sc i\kern-.025em b}\kern-.08em
    T\kern-.1667em\lower.7ex\hbox{E}\kern-.125emX}}

\usepackage{algorithm}
\usepackage[noend]{algpseudocode}
\usepackage{algorithmicx}

\begin{document}
\title{Clutter Tracking using Variational Message Passing\\
\thanks{
This work is funded by the Thomas B. Thriges Foundation grant 7538-1806.}
}

\author{
    \IEEEauthorblockN{Anders Malthe Westerkam, Troels Pedersen
    }
    \IEEEauthorblockA{Aalborg University, Aalborg Denmark. Email: \{amw, troels\}@es.aau.dk}
}

\maketitle

\begin{abstract}
We propose a message passing algorithm for tracking of clutter signals in MIMO radar. The method exploits basis expansion to linearise the signal model, to enable mean field approach for tracking the posterior distribution of the clutter as it evolves across time, as well as the mean and precision of the clutter map. The method shows good estimation accuracy in simulations for a scenario that adhere to the statistical model used for derivation as well as one that does not. The complexity of the method is linear in both the amount of parameters chosen and the amount of data under consideration.
\end{abstract}


\section{Introduction}
In recent years the use of drones have increased both privately, commercially, and militarily which makes privacy and security concerns a pressing matter, \cite{Poitevin2017}. In particular drones penetrating restricted airspace, at e.g. airports, may be catastrophic. The task of detecting drones is difficult due to their low radar cross section (RCS), velocity and altitude, \cite{Poitevin2017,Gong2023,Quevedo2019}. The low altitude results in many reflections off stationary ground objects such as vegetation or buildings referred to as clutter \cite{Richards2014}. 
The clutter signal may mask the return signal of the drone, resulting in unreliable tracking and in the worst case missed detection. For this reason clutter tracking, modelling, and mitigation is of great interest. As the scene under surveillance is ever-changing, a deterministic model is both unattainable and unrealistic. Instead a stochastic process is often sought.

Classically, the clutter has been modelled as a stochastic process where each range cell has some reflectivity drawn from a parameterised distribution. Most models used a combination of physical considerations with a random variable to model the return signal at the radar. In the widely used Gamma model \cite{Barton1985}, the parameters are obtained from measurements of similar terrains along with the angle of incidence for the radiation, requiring site specific knowledge. Another approach is presented in \cite{Capraro2007} where a map of the area under surveillance by a stationary radar is used to aid in clutter mitigation by supplying the radar with good priors based on landscape features.

Some work has been done in expressing the whole signal in terms of a random field, see e.g., \cite{Luo2022,Szabo}. In \cite{Szabo} it was shown that the matched filtered return signal amounts to a convolution between the wave ambiguity function and the clutter field, hence the inherent ambiguity of radar return signals are contained in this description. In \cite{Luo2022}, the ocean surface is modelled as a superposition of random Stokes waves and the clutter signal is calculated using a pulse expansion over the sea waves, i.e., the field is assumed constant in some small region. While \cite{Luo2022} showed great agreement with sea clutter near shore the extension to land clutter is not straightforward. However, as both descriptions are poorly suited for a general implementation for estimation of an arbitrary clutter field, \cite{Szabo} considers the signal post matched filtering which may lead to a prohibitively large data set for, e.g., multiple input multiple output (MIMO) radar systems.

This paper presents a method for clutter tracking. The clutter is modelled as a random field defined on some set of orthonormal separable basis functions. The expansion enables an extension of our previous work \cite{Westerkam2023} in the form of an algorithm using message passing on a Bayesian network which is then derived and is shown to be able to learn the parameters of a trial clutter field. The performance is evaluated, by simulation on a $4\times 4$ MIMO radar platform operating using a time division multiplexing (TDM) transmission scheme.

\section{Signal model}
Consider a scene with no target illuminated by a monostatic MIMO radar with $N_T$ isotropic transmitting antennas and $N_R$ receivers. Transmitter $m$ emits a signal $\text{Re}\left\{u^{(m)}(t)e^{i\omega_c t}\right\}$, where $\omega_c=2\pi f_c$ is the angular carrier frequency, $i$ is the imaginary unit, and $u^{(m)}(t)$ is the complex baseband signal which are mutually orthogonal. Transmission of all $N_T$ signals are occurring periodically with some pulse repetition frequency (PRF). 

We model the clutter signal as a superposition of reflected signals (single bounce) according to the clutter map $C(\mathbf{r};t)\in \mathbb{C}$, with position vector $\mathbf{r}\in \mathbb{R}^2$. We assume the clutter map to be slowly moving such that it can be viewed as constant over each MIMO transmission, $C(\mathbf{r};t) = C_n(\mathbf{r})$ for $n\Delta t \leq t < (n+1)\Delta t$, where $\Delta t=1/\text{PRF}$, likewise it will be assumed that the Doppler shift is zero for the whole map. Considering a narrow-band model and a small enough aperture that impinging waves are plane, the signal at the $j$-th receiver reads:
\begin{multline}\label{eq:signal_model}
    y_n^{(j)}(t) = \overbrace{\sum_{m=1}^{N_T} \int_{\mathbb{R}^2} C_n(\mathbf{r}) A^{(j,m)}(\theta) u^{(m)}\left(t-\tau\right)e^{i\omega_c(t-\tau)} d\mathbf{r}}^{s^{(j)}_{c,n}(t)} \\
    + w^{(j)}_n(t)
\end{multline}
Here $s_{c,n}^{(j)}(t)$ is the clutter signal, $\tau = 2|\mathbf{r}|/c$ is the propagation delay, $A^{(j,m)}$ is the steering matrix elements, $w^{(j)}_n(t)$ is white circular symmetric Gaussian noise. The clutter field is conveniently expressed in terms of an orthonormal separable basis as
\begin{equation}\label{eq:clutter_expansion}
    C_n(\mathbf{r}) = \sum_{k,l} \gamma^{(k,l)}_{n}\psi^{(k,l)}(\theta,r), 
\end{equation}
with expansion coefficients $\gamma^{(k,l)}_{n}$ and basis functions $\psi^{(k,l)}(\theta,r)$ satisfying orthonormality as
\begin{equation}
    \langle\psi^{(k',l')}|\psi^{(k,l)}\rangle = \delta(k-k')\delta(l-l').
\end{equation}
Here, $\langle\cdot|\cdot\rangle$ denotes the bra-ket notation for inner products, and $\delta(\cdot)$ denotes the Kronecker delta function. Requiring separable basis, i.e., $\psi^{(k,l)}(\theta,r) = \psi'^{(k)}(\theta)\psi'^{(l)}(r)$ enables expansion of the steering matrix elements as, 
\begin{equation}\label{eq:steering_matrix_element_expansion}
    A^{(j,m)}(\theta) = \sum_k \alpha^{(k,j,m)}\psi'^{(k)}(\theta),
\end{equation}
and the complex baseband signal as, 
\begin{equation}\label{eq:baseband_expansion}
    u^{(m)}(t-\tau)e^{i\omega_c (t-\tau)} = \sum_l \beta^{(l,m)}(t)\psi'^{(l)}(r).
\end{equation}
With expansion coefficients $\alpha^{(k,j,m)}$ and $\beta^{(l,m)}(t)$. Inserting (\ref{eq:clutter_expansion}), (\ref{eq:steering_matrix_element_expansion}), (\ref{eq:baseband_expansion}) into (\ref{eq:signal_model}) and integrating yields,
\begin{equation}\label{eq:inner_product_signal_model}
    y_n^{(j)}(t) = \sum_{m=1}^{N_T}\langle\text{conj}(\boldsymbol{\alpha}^{(j,m)})|\text{conj}(\boldsymbol{\gamma}_n)|\boldsymbol{\beta}^{(m)}(t)\rangle + w_n^{(j)}(t),
\end{equation}
where $\boldsymbol{\alpha}^{(j,m)}\in \mathbb{C}^{N_b\times 1}$, $\boldsymbol{\beta}^{(m)}(t)\in\mathbb{C}^{N_b\times1}$, and $\boldsymbol{\gamma}_n\in\mathbb{C}^{N_b\times N_b}$, with $N_b$ being the number of terms kept for the expansion. $\text{Conj}(\cdot)$ denotes complex conjugation.

The signal $y^{(j)}_n(t)$ is sampled  by the receiver with some frequency, giving rise to $N_s$ samples for each receiver. The collected data is organised in a vector as $\mathbf{y}_n\in\mathbb{C}^{N_sN_R\times 1}$. Likewise, the coefficient vectors can be organised into matrices $\Bar{\Bar{\boldsymbol{\alpha}}}^{(m)}\in \mathbb{C}^{N_R\times N_b}$, and $\Bar{\Bar{\boldsymbol{\beta}}}^{(m)}\in\mathbb{C}^{N_s\times N_b}$. Using these definitions the inner product in equation~(\ref{eq:inner_product_signal_model}) can be expressed as a linear transform on $\boldsymbol{\Gamma}_n=\text{conj}(\text{Vec}(\boldsymbol{\gamma}_n))$ with $\text{Vec}(\cdot)$ denoting the column-wise vectorisation,
\begin{equation}\label{eq:linear_signal_model}
    \mathbf{y}_n = \overbrace{\mathbf{M}\boldsymbol{\Gamma}_n}^{\mathbf{S}_{c,n}} + \mathbf{w}_n, \phantom{m} \mathbf{M} = \sum_m \Bar{\Bar{\boldsymbol{\alpha}}}^{(m)} \otimes \Bar{\Bar{\boldsymbol{\beta}}}^{(m)}.
\end{equation}
Here, $\otimes$ denotes the Kronecker product. The noise $\mathbf{w}_n\in\mathbb{C}^{N_sN_R\times 1}$ is distributed as  $\mathbf{w}_n\sim N^C(\boldsymbol{0},\lambda_W\mathbf{I})$, where $N^C(\boldsymbol{\mu},\boldsymbol{\Lambda})$ denotes a complex circular symmetric gaussian, with mean $\boldsymbol{\mu}$ and precision $\boldsymbol{\Lambda}$.

\section{Variational message passing}\label{sec:Variational_Message_Passing´}
\begin{figure}[t]
    \centering
    \scalebox{0.78}{
    \includegraphics[width= 1 \columnwidth]{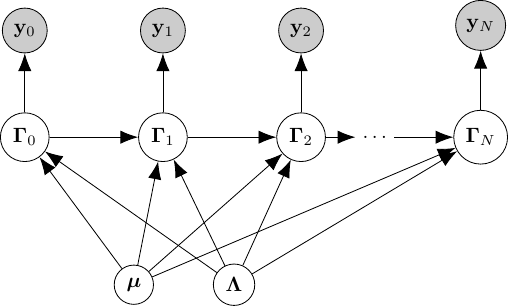}}
    \caption{Variational framework where the round nodes correspond to stochastic variables, and the shaded being observed variables}
    \label{fig:bay_network}    
\end{figure}

To infer a posterior distribution for $C_n(\mathbf{r})$ from the received data $\{\mathbf{y}_n\}$ we employ variational message passing \cite{bishop2007}. To that end, we impose a probability model in form of a Baysian network, Fig.~\ref{fig:bay_network}. We assume that the expansion coefficients are complex Gaussian,
\begin{equation}    \boldsymbol{\Gamma}_n|\boldsymbol{\mu},\boldsymbol{\Lambda} \sim N^C(\boldsymbol{\mu},\boldsymbol{\Lambda}).
\end{equation}
As such, the clutter tracking problem reduces to estimating $\{\boldsymbol{\Gamma}_n\}$, $\boldsymbol{\mu}$, and $\boldsymbol{\Lambda}$ from $\{\mathbf{y}_n\}$. Assuming that the coefficients are related by a Markov process results in the Bayesian network shown in Fig.~\ref{fig:bay_network} with joint probability distribution,
\begin{multline}\label{eq:joint_distribution}
    p(\{\mathbf{y}_n\},\{\boldsymbol{\Gamma}_n\},\boldsymbol{\mu},\boldsymbol{\Lambda}) =p(\mathbf{y}_0|\boldsymbol{\Gamma}_0)p(\boldsymbol{\Gamma}_N|\boldsymbol{\mu},\boldsymbol{\Lambda})p(\boldsymbol{\mu})p(\boldsymbol{\Lambda})\\ \times \prod_{q=0}^{N-1}\bigg[ p(\mathbf{y}_{N-q}|\boldsymbol{\Gamma}_{N-q})p(\boldsymbol{\Gamma}_{N-q}|\boldsymbol{\Gamma}_{N-1-q},\boldsymbol{\mu},\boldsymbol{\Lambda})\\\times p(\boldsymbol{\Gamma}_{N-1-q}|\boldsymbol{\mu},\boldsymbol{\Lambda})\bigg].
\end{multline}

By Bayes' theorem the posterior is proportional to the joint distribution \eqref{eq:joint_distribution}. Since marginalisation of this is intractable, we resort to a mean field approach \cite{bishop2007,Dauwels2007}, and approximate the distribution by a surrogate function as,
\begin{equation}
    q(\{\boldsymbol{\Gamma}_n\},\boldsymbol{\mu},\boldsymbol{\Lambda}) = q(\boldsymbol{\mu})q(\boldsymbol{\Lambda}) \prod_{q=0}^{N}q(\boldsymbol{\Gamma}_q).
\end{equation}
Minimising the Kullback-Leibler (KL) divergence with respect \eqref{eq:joint_distribution}, results in the following surrogate functions,
\begin{multline}\label{eq:max_KL_Gamma}
    \text{ln } q(\boldsymbol{\Gamma}_n) =\\ \text{ln }p(\mathbf{y}_n|\boldsymbol{\Gamma}_n) + \mathbb{E}_{\boldsymbol{\Gamma_{n-1}},\boldsymbol{\mu},\boldsymbol{\Lambda}}\left[\text{ln } p(\boldsymbol{\Gamma}_n|\boldsymbol{\Gamma}_{n-1},\boldsymbol{\mu},\boldsymbol{\Lambda})\right] \\+ \mathbb{E}_{\boldsymbol{\Gamma_{n+1}},\boldsymbol{\mu},\boldsymbol{\Lambda}}\left[\text{ln } p(\boldsymbol{\Gamma}_{n+1}|\boldsymbol{\Gamma}_{n},\boldsymbol{\mu},\boldsymbol{\Lambda})\right] \\+ \mathbb{E}_{\boldsymbol{\mu},\boldsymbol{\Lambda}}\left[\text{ln } p(\boldsymbol{\Gamma}_n|\boldsymbol{\mu},\boldsymbol{\Lambda})\right] +\text{const}, 
\end{multline}
\begin{multline}\label{eq:max_KL_mu}
    \text{ln } q(\boldsymbol{\mu}) = \text{ln } p(\boldsymbol{\mu}) + \sum_{n=0}^{N} \bigg[\mathbb{E}_{\boldsymbol{\Gamma}_n,\boldsymbol{\Lambda}}\left[\text{ln } p(\boldsymbol{\Gamma}_n|\boldsymbol{\mu},\boldsymbol{\Lambda})\right]\bigg] \\+ \sum_{n=1}^N\bigg[\mathbb{E}_{\boldsymbol{\Gamma}_{n},\boldsymbol{\Gamma}_{n-1},\boldsymbol{\Lambda}}\left[\text{ln } p(\boldsymbol{\Gamma}_{n}|\boldsymbol{\Gamma}_{n-1},\boldsymbol{\mu},\boldsymbol{\Lambda})\right]\bigg] +  \text{const},
\end{multline}
\begin{multline}\label{eq:max_KL_Lambda}
    \text{ln }q(\boldsymbol{\Lambda}) = \text{ln }p(\boldsymbol{\Lambda}) +\sum_{n=0}^{N}\bigg[\mathbb{E}_{\boldsymbol{\Gamma}_n,\boldsymbol{\mu}}\left[\text{ln } p(\boldsymbol{\Gamma}_n|\boldsymbol{\mu},\boldsymbol{\Lambda})\right] \bigg]\\+\sum_{n=1}^N\bigg[\mathbb{E}_{\boldsymbol{\Gamma}_{n},\boldsymbol{\Gamma}_{n-1},\boldsymbol{\mu}}\left[\text{ln } p(\boldsymbol{\Gamma}_{n}|\boldsymbol{\Gamma}_{n-1},\boldsymbol{\mu},\boldsymbol{\Lambda})\right] \bigg] + \text{const}.
\end{multline} 
Following \cite{Dauwels2007}, the posterior surrogates are determined by iterating through the terms in \eqref{eq:max_KL_Gamma}--\eqref{eq:max_KL_Lambda} i.e by message passing of the Bayesian network depicted in Fig.~\ref{fig:bay_network}. Hence to update the surrogate on a single node, a message is passed from all nodes in its neighbourhood, $\mathcal{N}_{(\cdot)}$. We calculate the messages as follows, starting with the message originating from the data,
\begin{equation}
    \text{ln } \varepsilon^{(\mathbf{y}_n \rightarrow \boldsymbol{\Gamma}_n)} = \text{ln } p(\mathbf{y}|\boldsymbol{\Gamma}_n),
\end{equation}
here $\varepsilon^{(\cdot)}$ denotes the message in question. By inspection of \eqref{eq:linear_signal_model}, $p(\mathbf{y}_n|\boldsymbol{\Gamma}_n) = N^C(\mathbf{M}\boldsymbol{\Gamma}_n,\lambda_W\mathbf{I})$. Unfortunately, due to the shape of $\mathbf{M}$, $p(\mathbf{y}_n|\boldsymbol{\Gamma}_n)$ is not guaranteed to be Gaussian in $\boldsymbol{\Gamma}_n$. Nonetheless, we enforce Gaussianity in both $\mathbf{y}_n$ and $\boldsymbol{\Gamma}_n$ for tractability so that,
\begin{equation}\label{eq:message_y_to_Gamma}
    \varepsilon^{(\mathbf{y}_n \rightarrow \boldsymbol{\Gamma}_n)} = N^C(\Bar{\varepsilon}^{(\mathbf{y}_n \rightarrow \boldsymbol{\Gamma}_n)},\left\{\Bar{\Bar{\varepsilon}}^{(\mathbf{y}_n\rightarrow\boldsymbol{\Gamma}_n)}\right\}^{-1}).
\end{equation}
Where the overbar designates the mean, and double overbar denotes the covariance matrix. By moment matching,
\begin{multline}
    \bar{\varepsilon}^{(\mathbf{y}_n \rightarrow \boldsymbol{\Gamma}_n)} = \int_{\boldsymbol{\Gamma}_n} \boldsymbol{\Gamma}_n p(\mathbf{y}|\boldsymbol{\Gamma}_n) d\boldsymbol{\Gamma}_n \\= \int_\mathbb{C}p(\mathbf{y}_n|\mathbf{S}_c)\int_{\boldsymbol{\Gamma}_n}\boldsymbol{\Gamma}_n\delta(\mathbf{S}_c-\mathbf{M}\boldsymbol{\Gamma}_n)d\boldsymbol{\Gamma}_n d\mathbf{S}_c.
\end{multline}
The innermost integral yields a response when $\mathbf{S}_c = \mathbf{M}\boldsymbol{\Gamma}_n$. This system of equations is over-determined, a solution is not guaranteed. However, an approximate solution in the 2--norm sense is obtained as $\boldsymbol{\Gamma}_n \approx \mathbf{M}^{\dagger}\mathbf{S}_c$, where $\mathbf{M}^{\dagger}$ is the Moore–Penrose pseudo inverse. Furthermore, the distribution $p(\mathbf{y}_n|\mathbf{S}_c)=p(\mathbf{y}_n-\mathbf{S}_c)$, and hence is Gaussian in both $\mathbf{y}_n$ and $\mathbf{S}_c$, the mean of the message can be found as,
\begin{equation}\label{eq:mean_of_message_from_y_to_Gamma}
    \Bar{\varepsilon}^{(\mathbf{y}_n \rightarrow \boldsymbol{\Gamma}_n)} = \int_\mathbb{C} \mathbf{M}^{\dagger}\mathbf{S}_cp(\mathbf{y}_n-\mathbf{S}_c) d\mathbf{S}_c = \mathbf{M}^{\dagger}\mathbf{y}_n.
\end{equation}
The same procedure gives the covariance matrix
\begin{equation}\label{eq:Covariance_of_message_from_y}
    \Bar{\Bar{\varepsilon}}^{(\mathbf{y}_n \rightarrow \boldsymbol{\Gamma}_n)} = \lambda_W^{-1}\mathbf{M}^{\dagger}\mathbf(\mathbf{M}^{\dagger})^H.
\end{equation} 

To derive the $\boldsymbol{\Gamma}_{n-1} \rightarrow \boldsymbol{\Gamma}_n$ message,
\begin{equation}\label{eq:message_pre_calc_n-1_n}
    \text{ln } \varepsilon^{(\boldsymbol{\Gamma}_{n-1}\rightarrow\boldsymbol{\Gamma}_{n})} = \underset{\boldsymbol{\Gamma}_{n-1},\boldsymbol{\mu},\boldsymbol{\Lambda}}{\mathbb{E}}\left[\text{ln } p(\boldsymbol{\Gamma}_n|\boldsymbol{\Gamma}_{n-1},\boldsymbol{\mu},\boldsymbol{\Lambda})\right],
\end{equation}
we assume that, conditional on $\boldsymbol{\mu}$ and $\boldsymbol{\Lambda}$, $\{\boldsymbol{\Gamma}_n\}$ forms a Markov process as,
\begin{equation}\label{eq:Markov_Chain}
    \boldsymbol{\Gamma}_{n} = \alpha\boldsymbol{\Gamma}_{n-1} + \mathbf{V}_N,
\end{equation}
with $\mathbf{V}_n \sim N^{C}(\boldsymbol{\mu}_V,\boldsymbol{\Lambda}_V)$ being the process noise. Assuming wide sense stationarity of $\{\boldsymbol{\Gamma}_n\}|\boldsymbol{\mu},\boldsymbol{\Lambda}$ implies the following mean and precision matrix,
\begin{equation}
    \boldsymbol{\mu}_V = \boldsymbol{\mu}(1-\alpha), \phantom{mm} \boldsymbol{\Lambda}_V = \frac{\boldsymbol{\Lambda}}{1-\alpha^2}.
\end{equation}
Inserting into \eqref{eq:message_pre_calc_n-1_n},
\begin{equation}\label{eq:message_from_n-1_to_n}
    \varepsilon^{(\boldsymbol{\Gamma}_{n-1}\rightarrow\boldsymbol{\Gamma}_n)} = N^C(\Bar{\varepsilon}^{(\boldsymbol{\Gamma}_{n-1}\rightarrow\boldsymbol{\Gamma}_n)},\left[\Bar{\Bar{\varepsilon}}^{(\boldsymbol{\Gamma}_{n-1}\rightarrow\boldsymbol{\Gamma}_n)}\right]^{-1}),
\end{equation}
with
\begin{equation}
    \Bar{\varepsilon}^{\boldsymbol{\Gamma}_{n-1}\rightarrow\boldsymbol{\Gamma}_n} = \Bar{\boldsymbol{\mu}} +\alpha(\Bar{\boldsymbol{\Gamma}}_{n-1}-\Bar{\boldsymbol{\mu}}),
\end{equation}
\begin{equation}
    \Bar{\Bar{\varepsilon}}^{\boldsymbol{\Gamma}_{n-1}\rightarrow\boldsymbol{\Gamma}_n} = (1-\alpha)^2 \Bar{\boldsymbol{\Lambda}}^{-1}. 
\end{equation}

In the same vein, the message $\boldsymbol{\Gamma}_{n+1}\rightarrow\boldsymbol{\Gamma}_n$ is,
\begin{equation}\label{eq:message_from_n+1_to_n}
    \varepsilon^{(\boldsymbol{\Gamma}_{n+1}\rightarrow\boldsymbol{\Gamma}_n)} = N^C(\Bar{\varepsilon}^{(\boldsymbol{\Gamma}_{n+1}\rightarrow\boldsymbol{\Gamma}_n)},\left[\Bar{\Bar{\varepsilon}}^{(\boldsymbol{\Gamma}_{n+1}\rightarrow\boldsymbol{\Gamma}_n)}\right]^{-1}),
\end{equation}
with
\begin{equation}
    \Bar{\varepsilon}^{(\boldsymbol{\Gamma}_{n+1}\rightarrow\boldsymbol{\Gamma})} = \Bar{\boldsymbol{\mu}}+\frac{1}{\alpha}(\Bar{\boldsymbol{\Gamma}}_{n+1}-\Bar{\boldsymbol{\mu}}),
\end{equation}
\begin{equation}
    \Bar{\Bar{\varepsilon}}^{(\boldsymbol{\Gamma}_{n+1}\rightarrow\boldsymbol{\Gamma}_n)} = \frac{1-\alpha^2}{\alpha^2}\Bar{\boldsymbol{\Lambda}}^{-1}.
\end{equation}

This leaves the message from $\{\boldsymbol{\mu},\boldsymbol{\Lambda}\}$ i.e., the last term of \eqref{eq:max_KL_Gamma}, which can be readily expressed as 
\begin{equation}\label{eq:message_muLambda_to_Gamma}
    \varepsilon^{(\boldsymbol{\mu},\boldsymbol{\Lambda}\rightarrow\boldsymbol{\Gamma}_n)} = N^C(\Bar{\boldsymbol{\mu}},\Bar{\Lambda}).
\end{equation}
Hence all the messages into $\boldsymbol{\Gamma}_n$ are complex Gaussian. The surrogate at $\boldsymbol{\Gamma}_n$ is the product of these,
\begin{equation}\label{eq:combi_of_gaussians}
    N^C(\boldsymbol{\mu}_{tot},\boldsymbol{\Lambda}_{tot}) = \prod_{n=0}^N N^C(\boldsymbol{\mu}_n,\boldsymbol{\Lambda}_n),
\end{equation}
with
\begin{equation}
    \boldsymbol{\Lambda}_{tot} = \sum_{n=0}^N \boldsymbol{\Lambda}_n, \phantom{mmm} \boldsymbol{\mu}_{tot} = \boldsymbol{\Lambda}_{tot}^{-1}\sum_{n=0}^N \boldsymbol{\Lambda}_n\boldsymbol{\mu}_n.
\end{equation}
Using this along with \eqref{eq:message_y_to_Gamma},\eqref{eq:message_from_n-1_to_n},\eqref{eq:message_from_n+1_to_n}, and \eqref{eq:message_muLambda_to_Gamma} the surrogate at $\boldsymbol{\Gamma}_n$ may be expressed in terms of a mean vector and precision matrix. 

The messages going to $\boldsymbol{\mu}$, starting with the first term in \eqref{eq:max_KL_mu}, i.e., the message $\{\boldsymbol{\mu},\boldsymbol{\Gamma}_n\}\rightarrow\boldsymbol{\mu}$, as $p(\boldsymbol{\Gamma}_n|\boldsymbol{\mu},\boldsymbol{\Lambda})$ is symmetric in $\boldsymbol{\Gamma}_n$ and $\boldsymbol{\mu}$ the form is the same as in \eqref{eq:message_muLambda_to_Gamma}, i.e.,
\begin{equation}
    \varepsilon^{(\{\boldsymbol{\Gamma}_n,\boldsymbol{\Lambda}\}\rightarrow\boldsymbol{\mu})}=N^C(\Bar{\boldsymbol{\Gamma}}_n,\Bar{\boldsymbol{\Lambda}}).
\end{equation}

The second term of $\eqref{eq:max_KL_mu}$, can also be shown to be Gaussian yielding the following,
\begin{equation}
    \varepsilon^{(\{\boldsymbol{\Gamma}_n,\boldsymbol{\Gamma}_{n-1},\boldsymbol{\Lambda}\}\rightarrow\boldsymbol{\mu})} = N^C\left(\frac{\left(\Bar{\boldsymbol{\Gamma}}_n-\alpha\Bar{\boldsymbol{\Gamma}}_{n-1}\right)}{1-\alpha},\frac{(1-\alpha)^2\Bar{\boldsymbol{\Lambda}}}{1-\alpha^2}\right).
\end{equation}
Noting that both messages going to $\boldsymbol{\mu}$ has constant precision matrix allows for a nice closed form of the surrogate in the $n$ messages as,
\begin{equation}\label{eq:postior_on_mu}
    q(\boldsymbol{\mu}) = N^C(\Bar{\boldsymbol{\mu}},\Bar{\Bar{\boldsymbol{\mu}}}^{-1}),
\end{equation}
with,
\begin{equation}\label{eq:covar_of_mu}
    \Bar{\Bar{\boldsymbol{\mu}}}^{-1} = \underbrace{\left[N+1 + N \frac{(1-\alpha)^2}{1-\alpha^2}\right]}_\kappa\Bar{\boldsymbol{\Lambda}},
\end{equation}
\begin{equation}\label{eq:mean_of_mu}
    \Bar{\boldsymbol{\mu}} = \kappa^{-1}\left[\sum_{n=0}^N \Bar{\boldsymbol{\Gamma}}_n + \frac{1-\alpha}{1-\alpha^2}\sum_{n=1}^N(\Bar{\boldsymbol{\Gamma}}_n-\alpha\Bar{\boldsymbol{\Gamma}}_{n-1})\right].
\end{equation}

Now to evaluate the terms in \eqref{eq:max_KL_Lambda}, for the marginal on $\boldsymbol{\Lambda}$, imposing a diagonal prior on $\mathbf{\Lambda}$, $p(\Lambda_{i\neq j})=0$, the expectation reads,
\begin{equation}
\mathbb{E}_{\boldsymbol{\Gamma}_n,\boldsymbol{\mu}}\left[p(\boldsymbol{\Gamma}_n|\boldsymbol{\mu},\boldsymbol{\Lambda})\right] = -\sum_{j=1}^{N_b}\mathbb{V}^{(n)}_{j}\lambda_j + \text{ln}(\lambda_j) + \text{const}. 
\end{equation}
\vspace{-8pt}with,
\begin{equation}\label{eq:V_def}
    \mathbb{V}^{(n)}_j = ||\Bar{\boldsymbol{\Gamma}}^{(n)}_{j}-\Bar{\boldsymbol{\mu}}_j||^2 + \Bar{\Bar{\boldsymbol{\Gamma}}}^{(n)}_{j,j} + \Bar{\Bar{\boldsymbol{\mu}}}_{j,j}.    
\end{equation}
Where the superscript denotes the time index. It appears that the message factorises in $\lambda_j$, similarly the second term of \eqref{eq:max_KL_Lambda} is of the form,\vspace{-8pt}
\begin{multline}
     \mathbb{E}_{\boldsymbol{\Gamma}_n,\boldsymbol{\Gamma}_{n-1},\boldsymbol{\mu}}\left[\text{ln}\left(p(\boldsymbol{\Gamma}_n|\boldsymbol{\Gamma}_{n-1},\boldsymbol{\mu},\boldsymbol{\Lambda})\right)\right] \\= -\sum_{j=1}^{N_b} \mathbb{W}_{j}^{(n,n-1)}\lambda_j + \text{ln}(\lambda_j) + \text{const},
\end{multline}
\vspace{-5pt} with,
\begin{multline}\label{eq:W_def}
     \boldsymbol{\mathbb{W}}^{(n,n-1)}_j = \frac{1}{1-\alpha^2}\bigg[||\Bar{\boldsymbol{\Gamma}}^{(n)}_j-\alpha\Bar{\boldsymbol{\Gamma}}^{(n-1)}_j-(1-\alpha)\Bar{\boldsymbol{\mu}}||^2 \\+ \Bar{\Bar{\boldsymbol{\Gamma}}}^{(n)}_{j,j} +\alpha^2\Bar{\Bar{\boldsymbol{\Gamma}}}^{(n-1)}_{j,j} + (1-\alpha)^2\Bar{\Bar{\boldsymbol{\mu}}}_{j,j} \bigg].
\end{multline}
Carrying out the summations and exponating the left hand side of \eqref{eq:max_KL_Lambda} the functional form becomes
\begin{equation}\label{eq:functional_form_of_q_Lambda}
    q(\lambda_j) \propto \lambda_j^{2N+1}e^{-(\sum_{n=0}^N \mathbb{V}_{j}^{(n)} + \sum_{n=1}^{N}\mathbb{W}_{j}^{(n,n-1)})\lambda_j},
\end{equation}
which is a gamma distribution, $\text{Gamma}(\zeta,\xi)$ with parameters $\zeta = 2N + 2$, and $\xi = \sum_{n=0}^N \mathbb{V}_{j}^{(n)} + \sum_{n=1}^{N}\mathbb{W}_{j}^{(n,n-1)}$.

It remains to consider the parameter $\alpha$. Principally, one could introduce $\alpha$ as a stochastic variable in the framework we forgo this step as it complicates the message structure. For simplicity however, we rely on a simple point estimate using the Yule-Walker approach.
\section{Algorithm}
The messages derived in Sec.~\ref{sec:Variational_Message_Passing´} enables formulation of the message passing algorithm. To this end we introduce an initialisation step, with a few further simplifications discussed below. The covariance $\Bar{\Bar{\varepsilon}}^{(\mathbf{y}_n\rightarrow\boldsymbol{\Gamma})}$ can be pre-calculated using \eqref{eq:Covariance_of_message_from_y} as it is unchanged during iterations. Furthermore, as the covariance is diagonally dominated we reduce complexity by only keeping the diagonal. The mean, $\Bar{\varepsilon}^{(\mathbf{y}_n\rightarrow\boldsymbol{\Gamma})}$ is calculated using \eqref{eq:mean_of_message_from_y_to_Gamma}, and $\{q(\boldsymbol{\Gamma}_n)\}$ are initialised by $\{\varepsilon^{(\mathbf{y}_n\rightarrow\boldsymbol{\Gamma}_n)}\}$. The mean, $\Bar{\boldsymbol{\mu}}$, represents the mean of the clutter field and as such is initialised by the sample mean of $\{\Bar{\boldsymbol{\Gamma}}_n\}$. The parameters for $q(\boldsymbol{\Lambda})$ may be initialised based on number of collected data points, as well as $\mathbb{V}^{(n}$ omitting $\Bar{\Bar{\boldsymbol{\mu}}}$ to avoid having to initialise this, and lastly the mean of $\boldsymbol{\Lambda}$ can be initialised in accordance with the underlying gamma distribution.
Lastly, $\alpha$ is initialised by solving the Yule-Walker equations for the set $\{\boldsymbol{\Gamma}_n\}$. 

The main algorithm consists of iterating through all data samples updating $q(\boldsymbol{\Gamma}_n)$ along the way, then updating $q(\boldsymbol{\mu})$ and $q(\boldsymbol{\Lambda})$, and optionally $\alpha$. This is done until convergence. The criterion of convergence could be set based on one of the parameters in the network, e.g, $\boldsymbol{\Lambda}$ as it was found it converges the slowest. However, in this article we fix the number of iterations to $N_I = 150$. The full algorithm can be seen in Alg.~\ref{al:algorithm}.

\begin{algorithm}[t]
\caption{Clutter Tracking}\label{algo:VMP}
\begin{algorithmic}[1]
\Procedure{VMP}{$\{\mathbf{y}_n\},\mathbf{M}$}
\Statex Initialisation:
\State Pre-calculate $\Bar{\Bar{\varepsilon}}^{(\mathbf{y}_n \rightarrow \boldsymbol{\Gamma}_n)}$ and $(\Bar{\Bar{\varepsilon}}^{(\mathbf{y}_n \rightarrow \boldsymbol{\Gamma}_n)})^{-1}$ using \eqref{eq:Covariance_of_message_from_y} 
\For{$n \leftarrow 0\,\, \text{to}\,\, N$}
\State $\Bar{\varepsilon}^{(\mathbf{y}_n \rightarrow \boldsymbol{\Gamma}_n)} \leftarrow \mathbf{M}^{\dagger}\mathbf{y}_n$, \eqref{eq:mean_of_message_from_y_to_Gamma}
\EndFor

\State $\Bar{\boldsymbol{\Gamma}}_n \leftarrow \Bar{\varepsilon}^{(\mathbf{y}_n 
\rightarrow \boldsymbol{\Gamma}_n)}$  
\State $\Bar{\Bar{\boldsymbol{\Gamma}}}_n \leftarrow \Bar{\Bar{\varepsilon}}^{(\mathbf{y}_n \rightarrow \boldsymbol{\Gamma}_n)}$
\State $\Bar{\boldsymbol{\mu}} \leftarrow \frac{1}{N+1}\sum_{n=0}^{N} \Bar{\boldsymbol{\Gamma}}_n$
\State $\zeta = 2N+2$
\State $\xi_j \leftarrow \sum_{n=0}^N |\Bar{\boldsymbol{\Gamma}}^{n}_j - \Bar{\boldsymbol{\mu}}_j|^2 + \Bar{\Bar{\boldsymbol{\Gamma}}}_{j,j}^{(n)}$
\State $\Bar{\boldsymbol{\Lambda}}^{-1} \leftarrow \frac{\boldsymbol{\xi}}{\zeta}$

\State Estimate $\alpha$ using Yule-Walker
\Statex Message Passing:
\For{$N_I$ iterations}
\For{$n \leftarrow 0\,\, \text{to}\,\, N$}  
\State $\Bar{\Bar{\boldsymbol{\Gamma}}}_n \leftarrow \sum_{\varepsilon\in\mathcal{N}_{\boldsymbol{\Gamma}_n}} \Bar{\Bar{\boldsymbol{\varepsilon}}}^{-1}$ 
\State $\Bar{\boldsymbol{\Gamma}}_n \leftarrow \Bar{\Bar{\boldsymbol{\Gamma}}}_n\sum_{\varepsilon\in\mathcal{N}_{\boldsymbol{\Gamma}_n}}\Bar{\Bar{\boldsymbol{\varepsilon}}}^{-1}\Bar{\boldsymbol{\varepsilon}}$
\EndFor
\State Update $\Bar{\boldsymbol{\mu}}$ using \eqref{eq:mean_of_mu} and $\Bar{\Bar{\boldsymbol{\mu}}}$ using \eqref{eq:covar_of_mu}
\State $\boldsymbol{\xi} \leftarrow \sum_{n=0}^N \boldsymbol{\mathbb{V}}^{(n)} + \sum_{n=1}^N \boldsymbol{\mathbb{W}}^{(n,n-1)}$,
\Statex \phantom{mmm,} with \eqref{eq:V_def}, and \eqref{eq:W_def}
\State $\Bar{\boldsymbol{\Lambda}}^{-1} \leftarrow \boldsymbol{\xi}/\zeta$,
\State (Optional) Update $\alpha$ using Yule-Walker
\EndFor
\EndProcedure
\end{algorithmic}
\label{al:algorithm}
\end{algorithm}
\section{Simulation}
\begin{figure}[tb]
\centerline{\includegraphics[width = 0.94\columnwidth]{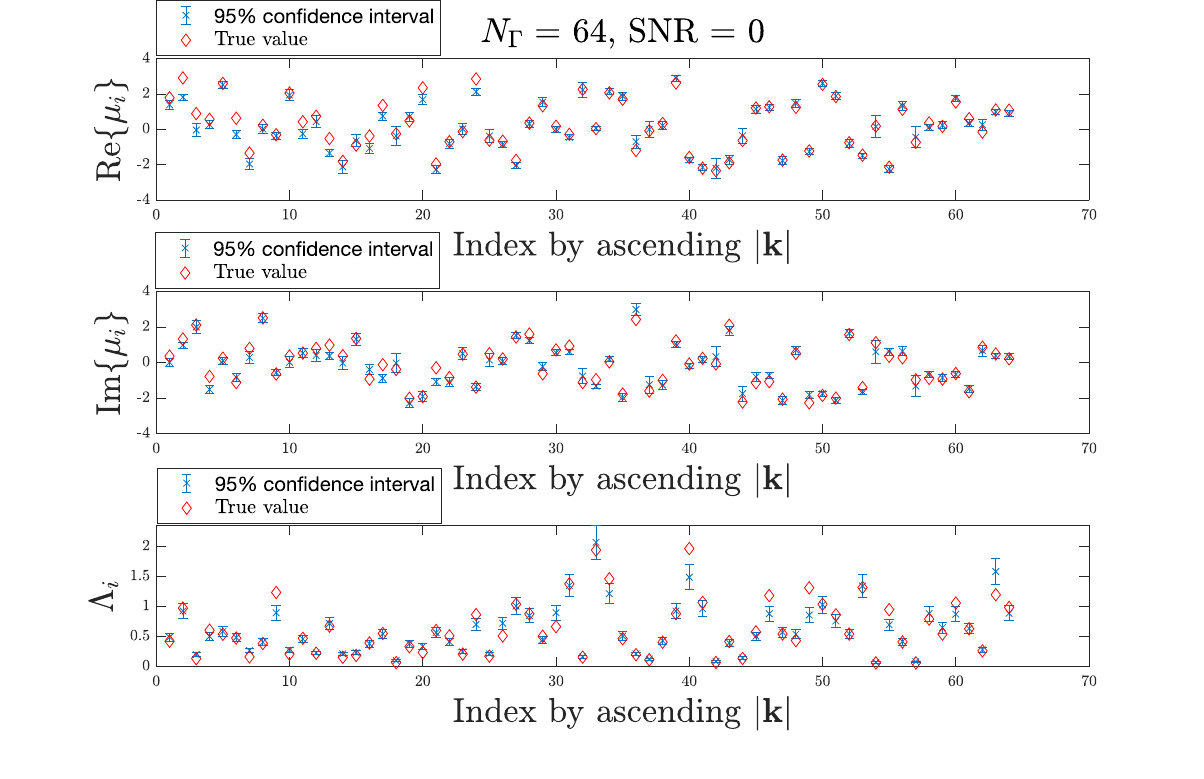}}
\caption{Scenario A. Estimated confidence intervals along with the real and imaginary part of the mean, as well as the elements of the precision matrix. Entries are sorted by ascending wave number hence the first entry is the DC component. The estimate of $\alpha$ settled at $\hat{\alpha}=0.169$. The radar uses the following parameters, PRF=10 Hz, Gain=1, $R_{max}=50$ m, $f_c = 10$ GHz, BW=$20$ MHz $T_{Tx}=16 \, \mu $s, $f_s=256$ MHz.}
\label{fig:est_acc}
\end{figure}

Two scenarios are considered:, (A) a "toy" scenario that exactly follow the Bayesian network depicted in Fig.~\ref{fig:bay_network}, (B) a clutter scene meant to resemble that observed in an experimental setting \cite{Lehmann2022}. For both scenarios we consider a $4\times 4$ MIMO radar operating in TDM mode, transmitting linear chirps of duration $T_{T_x}$, and sampling frequency $f_s$. The parameters of the radar for the purpose of simulation can be seen in the caption of Fig.~\ref{fig:est_acc}. All simulations are made in Matlab.

Scenario (A) is made to have the exact same statistical structure as considered when developing the algorithm, hence considering Fig.~\ref{fig:bay_network}. Thus we generate $\{\boldsymbol{\Gamma}_n\}$ in accordance with \eqref{eq:Markov_Chain} conditioned on $\boldsymbol{\mu}$ and $\boldsymbol{\Lambda}$.
Then each $\boldsymbol{\Gamma}_n$ is converted to $\boldsymbol{y}_n$ in acordance with \eqref{eq:linear_signal_model}, and the algorithm is run, we chose $\alpha = 0.1$, $N=99$ and to run it for $N_I = 150$ iterations. The result of running the algorithm on this toy scenario with a signal to noise ratio of 0 dB can be seen in Fig. \ref{fig:est_acc}. From here it is evident that the algorithm performs well even in low signal to noise ratios, with good estimation accuracy both in estimating the mean as well as the precision.

\begin{figure}
\begin{minipage}{.22\textwidth}
\centerline{\includegraphics[width = 1\textwidth]{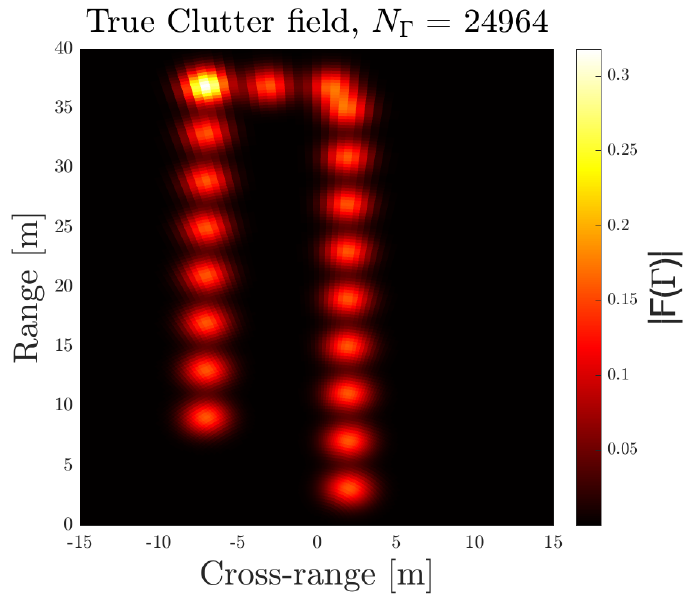}}
\centerline{(a)}
\end{minipage}
\hfill
\begin{minipage}{.22\textwidth}
\centerline{\includegraphics[width = 1\textwidth]{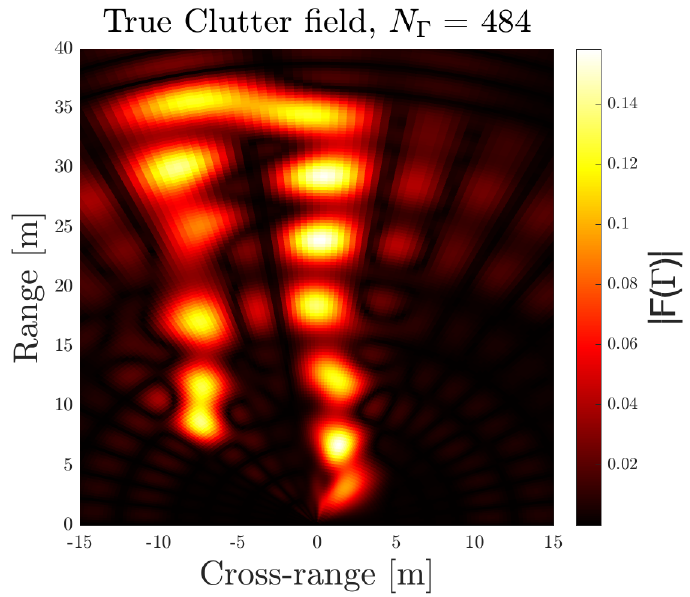}}    
\centerline{(b)}
\end{minipage}
\vfill
\begin{minipage}{.22\textwidth}
\centerline{\includegraphics[width = 1\textwidth]{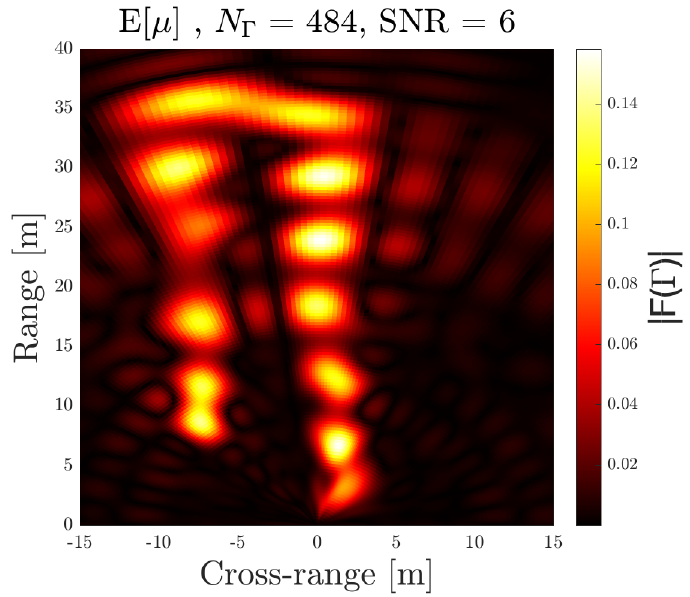}}    
\centerline{(c)}
\end{minipage}
\hfill
\begin{minipage}{.22\textwidth}
\centerline{\includegraphics[width = 1\textwidth]{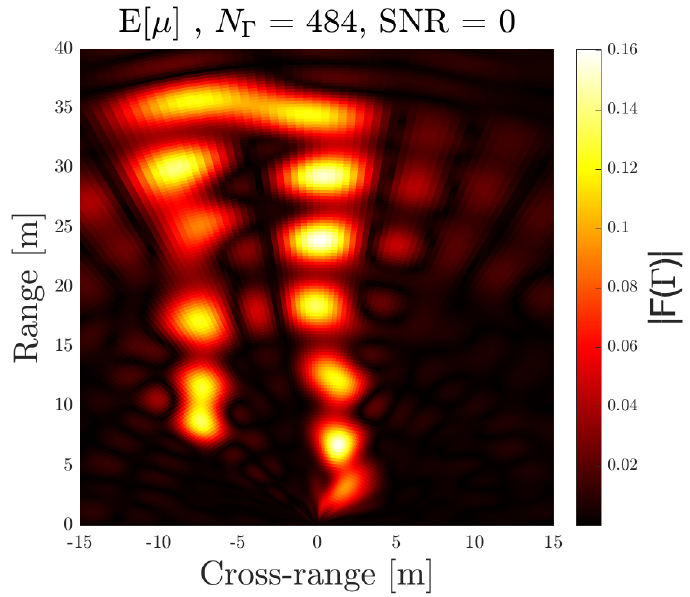}}    
\centerline{(d)}
\end{minipage}
\vfill
\begin{minipage}{.22\textwidth}
\centerline{\includegraphics[width = 1\textwidth]{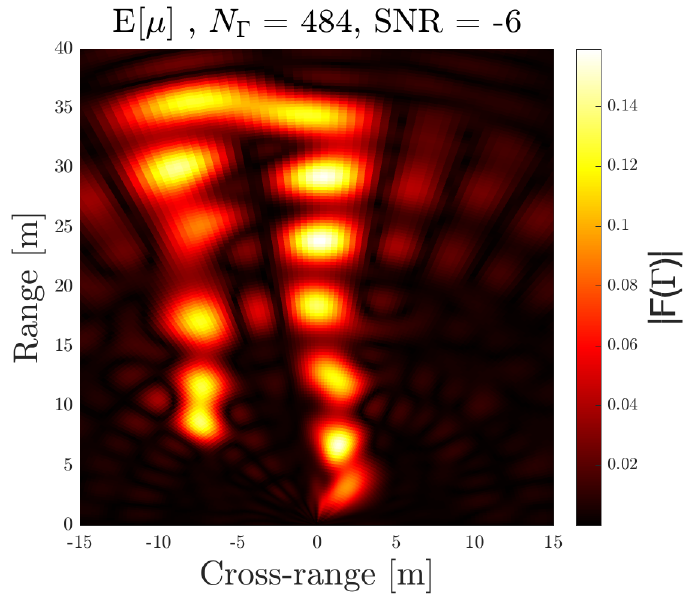}}    
\centerline{(e)}
\end{minipage}
\hfill
\begin{minipage}{.22\textwidth}
\centerline{\includegraphics[width = 1\textwidth]{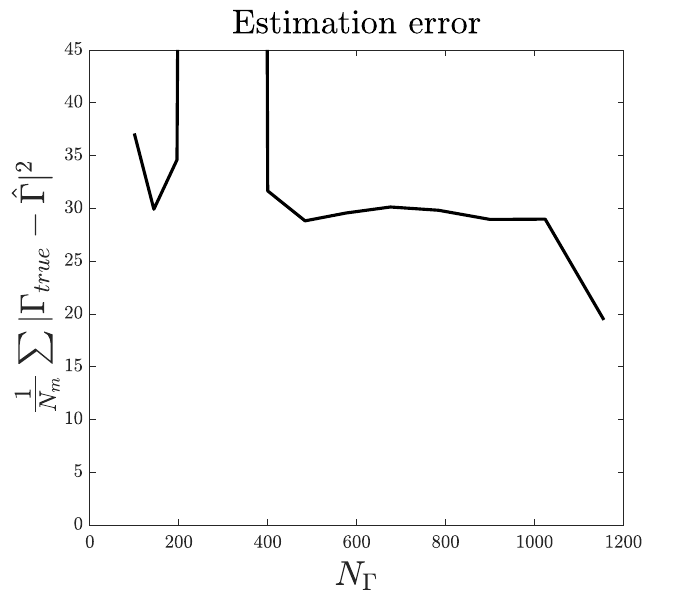}}    
\centerline{(f)}
\end{minipage}
\caption{Scenario B. \textbf{(a)} the real clutter map of the scene. \textbf{(b)} the clutter map of the scene as represented by choosing 484 coefficients. \textbf{(c)} the expectation of the $\boldsymbol{\mu}$ for the algorithm run on data with a SNR of 6 dB. Here the estimate of $\alpha$ settled at $\hat{\alpha}=0.85$ \textbf{(d)} the expectation of the $\boldsymbol{\mu}$ for the algorithm run on data with a SNR of 0 dB. Here the estimate of $\alpha$ settled at $\hat{\alpha}=0.82$  \textbf{(e)} the expectation of the $\boldsymbol{\mu}$ for the algorithm run on data with a SNR of -6 dB. Here the estimate of $\alpha$ settled at $\hat{\alpha}=0.72$ \textbf{(f)} Total estimation error measured by number of expansion coefficient, averaged over 100 Monte Carlo runs.}
\label{fig:Realizations_plot}
\end{figure}

For scenario (B) the map is made to resemble a fence surrounding a roof in the experimental setting in \cite{Lehmann2022} with a number of scatters placed to emulate fence posts. In simulation the evolution of $\boldsymbol{\Gamma}$ is still made to follow an auto--regressive process, however with a high precision to ensure that the map is initialised close to the mean, likewise $\alpha$ is chosen to be 0.9 to represent the slow variation of this static clutter map. 
Tests where performed weighing the number of coefficients with the estimation accuracy, Fig.~\ref{fig:Realizations_plot} (f), and a good compromise was found by using 484 coefficients. The representation of the true clutter map using this number of coefficients is shown in Fig.~\ref{fig:Realizations_plot} (b), while (c)-(e) is plotted using $\Bar{\boldsymbol{\mu}}$ estimated in differing noise scenarios. The algorithm captures the static clutter very well while filtering out the thermal noise only small deviations in the area surrounding the main scatters can be seen as the noise power is increased. Likewise the amplitude of the clutter map is preserved by the algorithm for the chosen number of coefficients. The coefficients in this map is also highly correlated and the ability of this framework hence suggest that the diagonalisation of $\Bar{\Bar{\varepsilon}}^{(\mathbf{y}_n\rightarrow\boldsymbol{\Gamma}_n)}$ as well as $\boldsymbol{\Lambda}$, does not significantly degrade the  ability to estimate the clutter map. Denoting the number of coefficients $N_\Gamma$, the complexity of the initialisation is $\mathcal{O}(N_RN_TN_sN_\Gamma^2)$ due to the matrix multiplication in \eqref{eq:Covariance_of_message_from_y}, however it can be done offline. The complexity of the algorithm proper is $\mathcal{O}(N_INN_\Gamma)$ with $N_I$ being the number of iterations until convergence. The algorithm it self is thus linear in both the number of parameters as well as the number of data frames.

\section{Conclusion}
The proposed Bayesian framework builds on orthogonal basis expansion to linearise the signal model, and reduce the number of parameters of the clutter signal across the whole field of view. The algorithm is based on the variational message passing framework; and results in low complexity, good estimation accuracy of the hyper parameters controlling the signal, posteriors of the network which are fully described by first and second order statistics, and good performance in low signal to noise ratios. This suggest a possibility of incorporating such an algorithm for clutter suppression for detection of weak targets such as drones.

\bibliographystyle{IEEEtran}
\bibliography{ES_lib.bib}

\end{document}